\documentclass{elsart}
\usepackage{epsfig}

\begin{document}

\begin{frontmatter}
\title{Rotational Reconstruction of Sapphire $(0001)$}
\author[Ljubljana]{Igor Vilfan,}
\author[Grenoble]{Fr\'ed\'eric Lan\c con}
\author[Grenoble]{and Jacques Villain}
\address[Ljubljana]{J. Stefan Institute, P.O. Box 3000, SI-1001 
      Ljubljana, Slovenia\\
e-mail: igor.vilfan@ijs.si}
\address[Grenoble]{D\'epartement de Recherche Fondamentale sur la 
      Mati\`ere Condens\'ee, CEA-Grenoble, F-38054 Grenoble cedex 9, France}

\begin{abstract}
The structure of the $(\sqrt{31}\times \sqrt{31})R\pm9^\circ$ reconstructed 
phase on sapphire (0001) surface is investigated by means of 
a simulation based on the energy minimization.
The interaction between Al adatoms is described with the semi-empirical
many-body Sutton-Chen potential, corrected for the charge transfer between
the metallic overlayer and the substrate. The interactions between the Al 
adatoms and sapphire substrate are described with a simple three-dimensional 
potential field which has the hexagonal periodicity of sapphire surface.
Our energy analysis gave evidence that the structure which is observed 
at room temperature is in fact a frozen  high-temperature structure.
In accordance with the X-ray scattering, a hexagonal domain 
pattern separated by domain walls has been found. 
The Al adatoms, distributed in two monolayers,  
are ordered and isomorphic to metallic Al(111) in the domains and 
disordered in the domain walls.  
The main reason for the rotational reconstruction is the lattice misfit
between the metallic Al and sapphire.

\end{abstract}



\end{frontmatter}

\section{Introduction}
Sapphire $\alpha - $Al$_2$O$_3$ is a technologically important material 
with a variety of $(0001)$ surface reconstructions.
Upon heating in UHV, the nonreconstructed $(1 \times 1)$ surface is stable 
up to $\sim$ 1250 C. Above this temperature, oxygen gradually 
evaporates 
whereas Al stays on the surface \cite{FS70}. As a consequence, the 
surface 
first reconstructs to $(\sqrt{3} \times \sqrt{3})$ rotated by 30$^\circ$
(or $(2\times 2)$, as reported by Gautier et al \cite{GDPVG91}), 
then to $(3\sqrt{3} \times 3\sqrt{3})$ rotated by 30$^\circ$ and 
finally,
after $\sim$ 20 minutes at 1350 C, to the $(\sqrt{31} \times \sqrt{31})
R\pm 9^\circ$ structure \cite{FS70,RVVB94}. 
It is this last reconstruction which will be the subject of the present 
paper.
The LEED pattern was first investigated by French and Somorjai \cite{FS70}
who found a cubic Al-rich surface structure over the hexagonal-symmetry 
bulk.
More recently, Vermeersch et al. \cite{VSLC90} obtained the same reconstruction 
after evaporating up to $\sim 2$ ML of Al on clean non-reconstructed 
sapphire surface. Gautier et al. \cite{GDPVG91} measured the electron 
energy loss spectra
and found a small hump in the sapphire band gap region. This hump is an
indication of metallic character of the Al-rich surface layer.
Renaud et al. \cite{RVVB94} made very precise grazing-incidence
X-ray diffraction (GIXD) measurements and were the first to propose a 
possible 
atomic structure of the $(\sqrt{31} \times \sqrt{31})$ structure. 
However, the Fourier transform of the GIXD provides the Patterson map and
not the real-space atomic structures. The lateral positions of Al
atoms in the disordered regions could not be uniquely determined.
In addition, the GIXD measurements could not tell the altitude of
individual atoms above the substrate.
Nevertheless, Renaud et al. \cite{RVVB94} found that 
the Al overlayer forms a hexagonal domain pattern.
Al atoms are ordered in two 
compact (111) planes of an FCC lattice, characteristic of metallic Al 
monolayers, within the domains 
and highly disordered in the ``domain walls'' between them. 
Whereas the origin of the domain pattern was attributed to the
lattice misfit between Al overlayer and the substrate, 
the reason for the $\pm 9^\circ$ rotation was less clear.

Interestingly enough, evaporation of two topmost planes of oxygens
from sapphire crystal leaves behind 5 Al atoms per surface unit cell, very 
close to the density 
observed by Renaud et al. \cite{RVVB94}. 

A mechanism of rotational reconstruction was proposed theoretically by 
Novaco and McTague \cite{NMcT77} and confirmed experimentally for 
several systems, including rare gases (physisorbed) 
and alkali metals (chemisorbed) on graphite \cite{SFC78,WHI91} and 
metals on metals \cite{TGS90,FD92}.
In the Novaco and McTague model, rotational reconstruction was 
attributed to  
lattice misfit combined with weak coupling between the overlayer and 
the substrate. This caused a weak
sinusoidal lateral distortion of the overlayer lattice. Interestingly 
enough, rotational reconstruction was observed for lattice misfits as 
large as 15 \% \cite{TGS90}.

Rotational reconstruction on sapphire, on the other hand, is different
and cannot be completely explained by the Novaco-McTague model. 
The origin of the 
reconstruction is still the lattice misfit between the overlayer and the
substrate, but now the distortion seems to be so strong that 
some atoms in the domain walls have lower coordination.
This strong disorder could be the consequence of strong
overlayer--substrate interactions and not of large lattice misfit. 
For the $(\sqrt{31}\times\sqrt{31})$ reconstruction the misfit between
the sapphire bulk and the Al metal overlayer is only $\sim 4$ \% and is
substantially less than in some other rotationally reconstructed 
systems \cite{TGS90}.
On the other hand, the interaction is strong also for some other 
rotationally 
reconstructed systems like Cs chemisorbed on graphite \cite{WHI91}.
There is another important difference
between the reconstructed sapphire and other systems: 
the overlayer on sapphire is composed of two monolayers \cite{RVVB94}
whereas in the other systems, it is composed of only one monolayer.
Evidently, the origin of rotational reconstruction and of strong 
disorder in the domain walls is not clear.  

In this paper we report on a simulation of the $(\sqrt{31}\times\sqrt{31})$
rotational reconstruction of sapphire with the aim of better 
understanding
the processes of rotational reconstruction on the sapphire and similar
surfaces and to help resolve some of the above open questions.

\section{The Model}

\begin{figure}
  \begin{center}
  \epsfxsize=8truecm
  \epsfbox{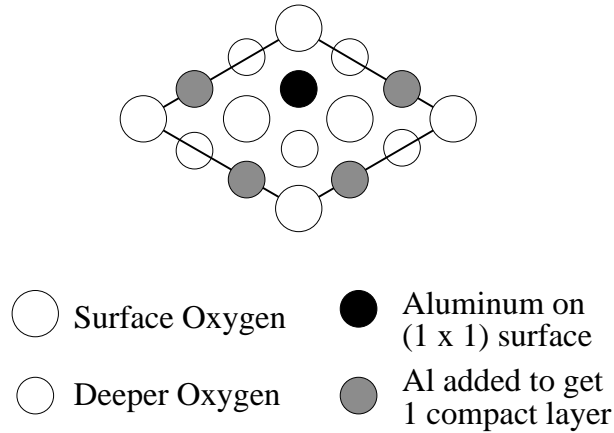}
  \end{center}
  \vskip5truemm
  \caption{Unit cell of non-reconstructed sapphire (0001).
      Al atoms (black circles) in the topmost plane are followed by
      an oxygen plane (open circles). Al atoms in the bulk of sapphire are 
      not displayed whereas the next oxygen layer is shown with smaller open 
      circles. Upon evaporation of oxygen, the density of surface Al atoms 
      increases. Shaded circles represent added Al atoms if sapphire were
       covered with one compact in-register Al monolayer.}  
  \label{nrsurf}
\end{figure}

The unit cell of clean, nonreconstructed sapphire (0001) surface is 
shown in Fig. \ref{nrsurf}. 
It is terminated with an Al plane with one Al atom per surface unit cell
(black disc; $1/3$ of a compact monolayer (ML) coverage)  \cite{TandMVG93}.  
Also shown 
(shaded discs) are Al atoms which have to be added to get one unexpanded 
compact monolayer, in register with sapphire.  

The reconstructed surface unit cell has about 157 Al atoms in the overlayer
\cite{RVVB94},
so one has to simulate rather large surface unit cells. 
This is possible only with efficient potentials,
therefore we describe the interaction between the overlayer atoms with 
the   
semi-empirical Sutton-Chen potential \cite{SC90,FS84} which 
has been also used in studying surface properties,
including reconstruction, of FCC metals \cite{TLB93}.
These potentials are many-body potentials 
with elements of two-body terms in it and are written in the form: 

\begin{equation}
   U = {1\over 2}\sum_{i\ne j} \epsilon \left( {a\over r_{ij}}\right)^n  - 
   \epsilon C \sum_{i} \sqrt{\rho_i}
   \label{Alpot}
\end{equation}
($r_{ij}$ is the separation between the atoms $i$ and $j$). 
The first term in (\ref{Alpot}) 
represents the core repulsion potential and the second term 
the bonding energy mediated by the electrons.  
$\rho_i$ is an effective local electron density at the site $i$
and is written as:
\begin{equation}
    \rho_i = \sum_{j\ne i}\left( {a\over r_{ij}}\right)^m.
\end{equation}
$\epsilon$ and $C$ are parameters of the model which, together 
with the exponents $n$ and $m$, determine the repulsive and cohesive
energies, respectively.   $a$ is the lattice constant of an Al FCC crystal.
Sutton and Chen \cite{SC90} published the following values for the 
potential parameters of Al: 
\begin{equation}
   m = 6, \quad n = 7, \quad \epsilon = 33.147\, \mbox{ meV},\quad 
   \epsilon C = 16.399\mbox{ meV}.
\end{equation}
We truncated the potential continuously (with a fifth order 
polynomial) between $r/r_0 = 3.17$ and 
$3.32$
($r_0$ is the nearest-neighbour distance). In this way the cutoff is
between the 10$^{\mbox{th}}$ and 11$^{\mbox{th}}$ shells of neighbours. 
That meant that interaction with 68  neighbours were included 
if Al were perfectly ordered in two FCC(111) planes.
With this set of parameters, the cohesive energy in the bulk Al 
is 3.313 eV/atom, compared to 3.34 eV/atom in \cite{SC90}.
The difference comes from different cutoff radii. 

The origin of bonding is in the electrons of the overlayer. 
On the reconstructed surface, there are
about 5 Al atoms per nonreconstructed unit cell, on the average.
Each Al atom has two $3s$ and one $3p$ electrons which all contribute to 
the cohesive energy.
Out of these 5 atoms, one is already present on the non-reconstructed 
surface and is ionically bound to the sapphire substrate.
This reduces the effective density of the electrons in the conduction 
band by $1/5$ and thus scales
the constant $\epsilon C$ in equation~(\ref{Alpot})
by a factor $\sqrt{4/5}$ (\textit{i.e.}, $\epsilon C = 14.668$ eV). 
$\epsilon C$ can be additionally reduced by the charge transfer 
between the metallic
overlayer and the insulating substrate~\cite{GDPVG91}. 
It is not known how strong the charge transfer is, therefore we treated
the constant $C$ as an adjustable parameter. 

The interaction between the overlayer and the substrate was modelled 
by a simple periodic potential.  
The substrate is much stiffer than the Al overlayer, therefore
the relaxation of the substrate caused by the overlayer was neglected 
in the simulations.
The substrate potential was expanded in a power series and 
only the six lowest-order
terms were retained, similarly as in the studies of noble gases 
on graphite, 
\begin{equation}
   U_S = {U_{L} - \cos (\vec{k}_1 \cdot \vec{r}) \cos(\vec{k}_2 \cdot 
\vec{r})
             \cos(\vec{k}_3 \cdot \vec{r}) \over U_{L} - 1}\; U_{LJ}(z).
\label{Usub}
\end{equation}             
\begin{equation}
        \vec{k}_1 = {2\pi \over a_s} (0,1),  \qquad
        \vec{k}_2 =  {\pi \over a_s} ( \sqrt{3}, -1 ),  \qquad
        \vec{k}_3 =  {\pi \over a_s} (-\sqrt{3}, -1 )  
\end{equation}
are the unit vectors in the plane of the surface, and $a_s$ the substrate 
lattice constant.
$U_{L}$ controls the lateral modulation of the potential, and $U_{LJ}$,
\begin{equation}
    U_{LJ}(z) = U_0 \left[ \left({z_0\over z}\right)^{12} - 
                         2 \left({z_0\over z}\right)^{6}\right],
\label{U_LJ}
\end{equation}
its $z-$dependence, in the direction perpendicular to the surface.
This dependence 
is necessary because the overlayer is more than one monolayer thick.
In the Lennard-Jones potential (\ref{U_LJ}),  
$U_0$ is the depth of the substrate potential and $z_0$ determines 
the position of the minimum and the width of the potential in the 
vertical direction.
Thus, the overlayer -- substrate potential is described by three 
variational parameters, $U_L$, $U_0$, and $z_0$. 
\begin{figure}
  \begin{center}
  \epsfxsize=8truecm
  \epsfbox{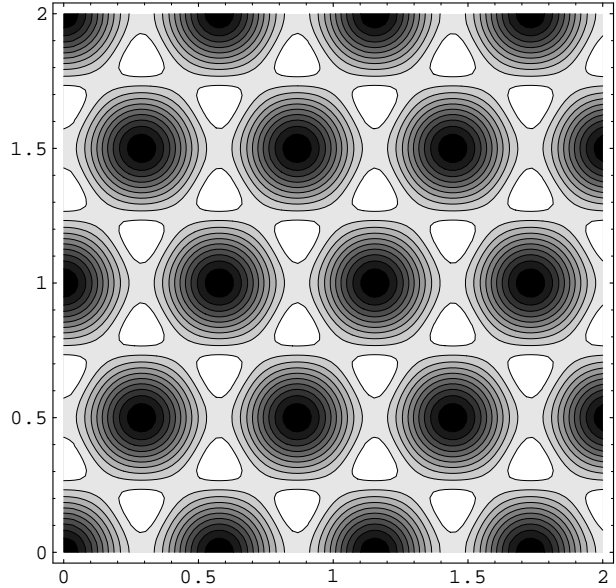}
  \end{center}
  \vskip5truemm
  \caption{The substrate potential has six equally deep minima (white) around 
   each maximum (black) which is located above the last layer of oxygens.}  
  \label{sub_pot}
\end{figure}
This potential, shown in Fig \ref{sub_pot}, has six equally deep minima 
around
each topmost oxygen atom of sapphire substrate. The potential (\ref{Usub})
is not
deeper on the sites, already occupied by Al atoms on the 
unreconstructed (0001)
surface. This simplification is backed by a careful inspection of 
the real-space overlayer structure obtained from the 
diffraction data  \cite{RVVB94}. During the process of reconstruction,
the original Al atoms also moved to other positions. 
They are not stronger bound than any 
other Al atoms coming to the surface during oxygen evaporation. 

We simulated a box with the base plane of 
$\sqrt{93}a \times \sqrt{31}a$.
Such a rectangle accommodates exactly  
two reconstructed surface unit cells under periodic boundary
conditions for the overlayer and substrate. 
The in-plane symmetry axis of the substrate 
is rotated with respect to the rectangle by 
$\alpha = \sin^{-1}(\sqrt{93}/62)\approx 8.9^\circ$.
In the initial configurations the Al atoms were put on two planes and
positioned close to the positions found by Renaud et al. \cite{RVVB94}. 
We started with this
configuration because of the problems with metastability.

Our results have been compared with the GIXD result of Renaud et al. 
\cite{RVVB94} who found 157 adatoms in a surface unit cell,
distributed in two monolayers. In their analysis, however, some 
atomic sites could have been only partially occupied. Therefore we
performed simulations with the number of Al atoms in the overlayer varying
from 302 to 318 adatoms (in two unit cells).

For a given set of  the potential parameters $C$, $U_L$, $U_0$ and $z_0$,
the equilibrium Al atomic positions were calculated by the energy minimization.
Once the atomic positions were determined, we calculated the corresponding 
diffraction pattern, given by the structure factors $S_c(\vec{Q})$ of the 
Bragg peaks at $\vec{Q} = (h,k)$, and compared it with the experimental 
$S_e(\vec{Q})$ \cite{RVVB94,Rpi}.
To evaluate the quality of the fit we first scaled $S_c(\vec{Q})$ with a
factor $s$ and then calculated the residue, introduced as
\begin{equation}
   \chi^2 = {\sum_Q [S_e(\vec{Q}) - s S_c(\vec{Q})]^2 
                        \over \sum_Q [S_e(\vec{Q})]^2}.
\end{equation}
The scaling factor $s$ was chosen to minimise the residue $\chi^2$:
\begin{equation}
   s = {\sum_Q [S_e(\vec{Q}) S_c(\vec{Q})] \over \sum_Q [S_c(\vec{Q})]^2}.
\end{equation}
Summation over the first $17 \times 16$ 
nonequivalent diffraction peaks in the reciprocal space was performed.
The parameters $C$, $U_L$, $U_0$ and $z_0$ were then varied to find 
the minimal $\chi^2$.

In the process of relaxation towards the (local) minimum energy 
configuration with the lowest residue, 
severe problems with hysteresis and local 
energy minima (metastability) were encountered. This is not too surprising
since the ``domain wall'' region is full of lattice defects.
To avoid trapping in
a metastable state, the relaxed structure was shaken randomly and relaxed 
again. 

\section{Results}

The best agreement between the calculated and experimental diffraction
patterns 
was found, and the most extensive simulations have been done for 
the structure with 314 adatoms (157 per reconstructed
surface unit cell).
\begin{figure}
  \begin{center}
  \epsfxsize=14truecm
  \epsfbox{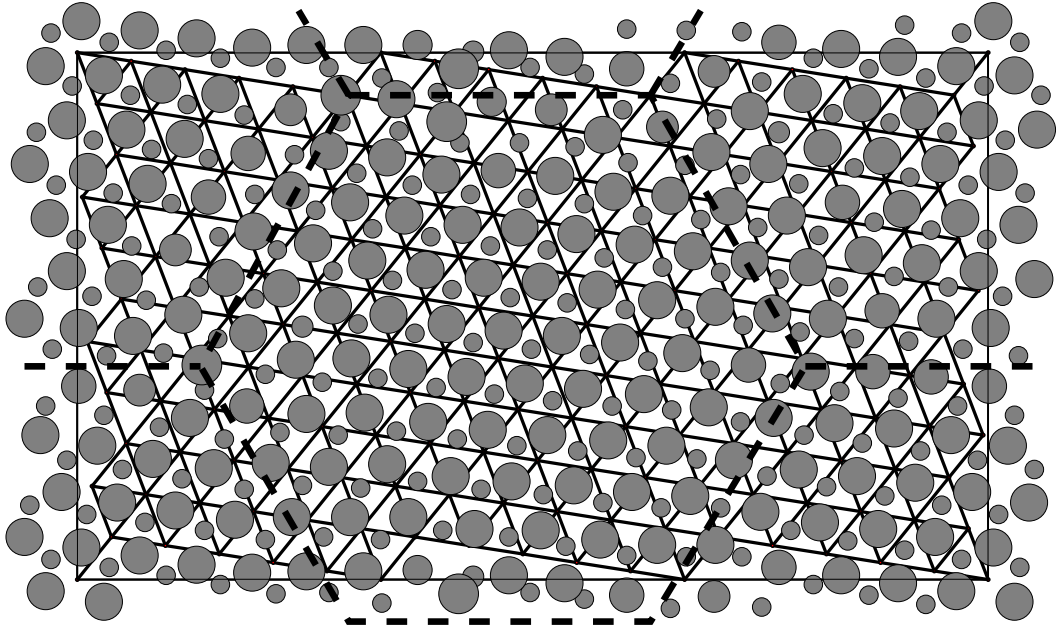}
  \end{center}
  \vskip5truemm
  \caption{Structure of the Al overlayer obtained in the simulation of 314
  atoms in a rectangle of the size $\sqrt{93} a \times \sqrt{31} a$ with
  periodic boundary conditions. The corners 
  of the underlying triangles show the positions of the topmost oxygens 
  of sapphire, where the substrate potential is maximal. The potential 
  is minimal in the centres of the triangles. 
  The grey circles represent the Al atoms and their radii are proportional
  to the altitude of the atom above the substrate. 
  The Al atoms are arranged in two monolayers, the larger circles 
  show the atoms in the upper layer and the smaller circles the atoms 
  of the lower layer. The atoms are more ordered and closer to the minima
  of the substrate potential 
  in the hexagonal domains and more disordered close to the 
  walls (dashed lines) separating the domains. Notice that the two 
  domains do not have the same structure in the disordered regions.}  
  \label{rec_struc}
\end{figure}
The best fit between with 314 atoms was obtained for
\begin{equation}
   \epsilon C = 13.87 \mbox{eV} \qquad
   U_L = 4.56          \qquad
   U_0 = 1.35 eV       \qquad
   z_0 = 2.64 \mbox{\AA}.
\end{equation}
The energy of this structure is 
$E = - 3.21$ eV/adatom. The average core repulsion energy between the 
adatoms is $1.65$ eV, the bonding energy between them is $-3.95$ eV, and 
the average interaction energy with the substrate is $-0.92$ eV.
 
The real-space structure with the best fit is seen in Fig. \ref{rec_struc}. 
Comparison of the real-space structures obtained from the simulation and 
from the
GIXD shows that the simulated structure is substantially more ordered
and that the simulated disordered regions, domain walls are
broader and less pronounced. 
The two domains in the simulated box do not have the same 
structure after relaxation although their initial configurations were
identical. This is an indication of many
metastable states in the disordered regions.

\begin{figure}
  \begin{center}
  \epsfxsize=12truecm
  \epsfbox{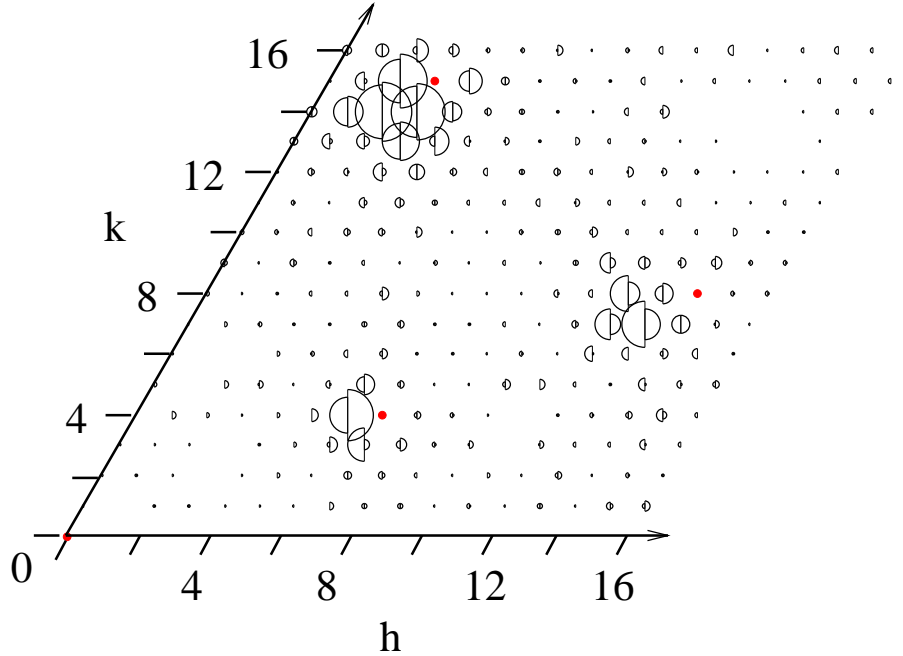}
  \end{center}
  \vskip5truemm
  \caption{Calculated (left-hand semicircles) and experimental 
  (right-hand semicircles) \cite{RVVB94} diffraction patterns, 
  indexed in the reciprocal space of the reconstructed surface unit cell. 
  Small black disks are the bulk allowed reflections.}  
  \label{dif_pat.eps}
\end{figure}

The residue can be compared with the 
experimental uncertainty $\sigma$, defined as
\begin{equation}
    \sigma^2 = {\sum_Q [\sigma(\vec{Q})]^2 
                        \over \sum_Q [S_e(\vec{Q})]^2},
\end{equation} 
where $\sigma(\vec{Q})$ are the experimental uncertainties of the 
structure factors at the Bragg points $\vec{Q}$. Using the experimental
data of Renaud \cite{Rpi}, we found $\sigma = 0.28$ whereas our best fit
had $\chi = 0.5$. 
Comparison of the calculated and experimental structure factors, 
Fig. \ref{dif_pat.eps} shows good agreement of 
the intense peaks around the substrate Bragg peaks.
Some discrepancy is observed for less intense peaks further away from the
substrate Bragg peaks. The biggest difference in the structure
factors is found for $\vec{Q} = (7,3)$. The reason is most probably in 
the  simple substrate potential used, so one cannot
expect to find exactly the same structure as on a real sapphire surface.

\section{Discussion}

The simulations were done by energy minimization.
In principle, however, the
energy minimization is acceptable only for low temperatures, when the 
entropy is small, and one should minimize the free energy. 
It is expected that the structure observed in the experiment at room
temperature has minimal free energy at the freezing 
temperature which is
substantially higher than the room temperature.  
At the freezing temperature, the contribution of the entropy is 
significant.
This -- together with the simplified overlayer-substrate potential -- 
possibly explains why the more disordered structure with the best 
fit has higher energy than
some other --  more ordered -- structures with worse fit. 
Indeed, the energy difference between different structures, 
simulated with the same
variational parameters, was of the
order $0.01$ eV/atom (if distributed uniformly among 314 atoms) and is of the 
same order as the contribution of the entropy 
to the free energy at the freezing temperature.

Another mechanism which would lead to a metastable overlayer structure
could be channeling, transport of Al atoms directly into a strongly
metastable state on the surface during the process of oxygen 
evaporation. However, it is very unlikely that the metastable potential 
minima would be so deep that 
diffusion at $\sim$1000 K couldn't bring the overlayer structure to an 
equilibrium.

Renaud et al. were not able to tell in their paper \cite{RVVB94} 
which Al plane is higher. 
They did, however, distinguish between the more ordered and 
more disordered planes. On the basis of preliminary numerical relaxation
they anticipated that the more disordered layer was the lower plane.
Our simulation does not support this distinction, we found both 
planes equally disordered. 

if the simulations were done with 314 Al atoms,  158 atoms were found
in the lower layer, which is much flatter than the upper layer. 
In the hexagonal domains, the Al atoms are well ordered in both planes
like in metallic Al(111). 
In the domain walls, the Al atoms are strongly disordered. In the 
simulation,
the disorder is manifested also in such a way that equivalent atoms 
in different reconstructed unit cells had 
different positions, the translational 
symmetry of the reconstructed unit cell was broken.  

Simulations with different number of Al atoms, $N_{Al}$, show a strong
increase in $\chi$ if  $N_{Al} > 314$ and a small increase if 
$N_{Al} < 314$. For $N_{Al} > 314$, Al atoms very often moved into 
the third plane.

In our ``best fit'' structure, Fig. \ref{rec_struc}, the disordered 
regions (domain walls) were 
less pronounced than in the structure proposed by Renaud et al. \cite{RVVB94}.
Our structure has a smoother transition between the domains, 
one could say that our domain walls are broader. 
It is not clear to which extent this difference 
is the consequence of approximate description of the substrate potential.

With the substrate potential described by Eq. (\ref{Usub}), 
one cannot study the process of oxygen evaporation, 
we cannot say whether two overlayers of Al are stable or only a 
transient state and more Al layers are formed upon further heating.
A stabilizing mechanism, discussed in detail by Chen et al. \cite{CLT91}, 
is based on the attraction between two surfaces across a thin metal, 
similar to the Casimir effect. 
This attraction is described by the Hamaker constant. 
We estimated the Hamaker constant of Al metal between sapphire insulator
and vacuum, $H \approx - 3.7 \times 10^{-2}$ eV. 
The corresponding energy gain is of the order $10^{-3}$ eV/(Al atom). 
We conclude that this mechanism of attraction between the two adjacent 
interfaces is too weak to stop the overlayer growth at two monolayers. 
In fact, the 
experimentalists observe that further heating in UHV brings more
Al atoms to the surface \cite{Rpi}, the two-layer 
$(\sqrt{31}\times \sqrt{31})R\pm9^\circ$ reconstructed structure is thus
not an equilibrium state of sapphire at 1350 C. 
Thus, in UHV, more Al atoms build next monolayers whereas heating 
under oxygen atmosphere oxidizes and deconstructs the surface. 

The fact that a metallic type potential, Eq.~(\ref{Alpot}), 
leads to a good model of this reconstructed sapphire surface
is consistent with the metallic character of its Al overlayer.
The minor discrepancy in some diffraction peaks is probably due to
the oversimplified substrate potential which cannot lead to the
exact details of the reconstruction. But the good overall agreement
of the diffraction patterns shows that we have used 
reasonable interatomic interaction ingredients to simulate this
reconstruction.

\ack{ The authors are indebted to G. Renaud for many interesting 
discussions and for providing them his experimental results. 
The financial support of the
French-Slovenian programme Proteus is deeply acknowledged.}

\end{document}